\documentclass[]{article}

\usepackage{amsmath}
\usepackage{amsfonts}
\usepackage{amssymb}
\usepackage{graphicx}
\usepackage{hyperref}
\usepackage{subcaption}
\usepackage{extarrows}
\usepackage{dirtree}
\usepackage{multirow}
\usepackage{mathtools}
\usepackage{tikz-cd} 
\usetikzlibrary{arrows.meta, positioning, calc, fit}
\usepackage{pst-node}
\usepackage{cleveref}
\usepackage{booktabs}
\usepackage{fullpage}
\usepackage{pgfplots}
\pgfplotsset{compat=1.18}
\usepackage{float}

\usepackage[style=numeric,backend=bibtex,sorting=none]{biblatex}
\title{Point Group Equivariant Graph Neural Networks for Materials}
\author{Alexander J. Heilman, Qimin Yan}

\begin{document}
	
	\maketitle
	
	\begin{abstract}
		Equivariant graph neural networks have proven effective tools for inference of material's properties directly from their structure. Traditionally, these have been applied such that they respect full $O(3)$ equivariance, so that any rotation or reflection of the input structure is respected in the model's output. While this works for general arrangements of atoms, additional symmetries of atomistic systems are left unleveraged. Furthermore, any symmetries of the filter functions are implicitly learned from the full dataset and not strictly enforced. In this work, we introduce point-group symmetry aware equivariant graph neural networks (PGEqNN) for materials science, with filter functions aligned with symmetry-aware indices for greater granularity in predictive tasks. With this architecture, we show that most of the predictive power of equivariant networks for tensorial elastic and dielectric datasets lies in the trivial subspaces of the point-group adapted bases. Exploiting this, an $A_1$-restricted variant matches or improves on its full point-group and $SO(3)$-partitioned counterparts while training fewer active parameters, yielding leaner models of equal accuracy.
	\end{abstract}
	
	\section*{Introduction \& Motivation}
	Machine learning has proven an effective paradigm for the prediction of material properties directly from their structure  \cite{mlreview1, mlreview2, mlreview3, cgcnn, chgcnn, megnet, m3gnet}. In particular, recent advancements have proven $SO(3)$ equivariant networks, that is, networks that respect rotations, to be exceptionally effective for use in physics \cite{e3nn, tfn, tensor_pred, gnnopt}. In addition to better absolute performance on traditional scalar learning tasks, equivariant networks allow for the direct prediction of directional quantities such as dipole moment and tensorial quantities such as elastic and dielectric response. However, while these equivariant networks can capture the symmetry of the material systems in their learned feature vectors, the models themselves often are agnostic to the underlying symmetries, leaving additional connections based upon these symmetries unleveraged in their learned weights. In this work, we present point group symmetry-aware equivariant networks, which can support full equivariance while simultaneously taking advantage of these higher-order connections supported by the symmetry of the systems under consideration. This is achieved by partitioning filter functions, weights, and features by not only their $SO(3)$ rotational order irreducible representations (IRs), but further their point group irreducible representations, allowing for specificity within rotational order not possible within traditional $SO(3)$ equivariant networks.
	
	Beyond introducing the architecture, we ask which of its ingredients actually carry the benefit. To this end we compare, at parameter counts matched within every rotational order, four models spanning the range of symmetry awareness: an invariant scalar baseline, the $SO(3)$ partition, the full point-group partition, and an $A_1$-restricted variant confined to the trivial-irrep blocks. Across four elastic and four dielectric datasets drawn from the Materials Project, the predictive content of the point-group partition concentrates in its trivial blocks: the $A_1$-restricted model matches or improves on both references with fewer active parameters, and the full partition pays off only where the data carry genuinely learnable anisotropic signal.
	
	A closely related strategy in computer vision is image \textit{canonicalization} \cite{kaba2023}. Rather than constraining the architecture to be equivariant by construction, a small auxiliary network is learned that maps each input to a single representative of its orbit under the symmetry group, after which a standard non-equivariant backbone handles the downstream prediction. This idea traces back to the spatial-transformer networks of \cite{stn}, in which a learned affine warp factors out scale, rotation, and small deformations before the recognition layers see the input. More recent work has reformulated the same intuition as an exact equivariant adapter: \cite{kaba2023} learn the canonicalization function end-to-end and report performance competitive with hard-equivariant baselines across image classification, point clouds, and N-body dynamics, while \cite{mondal2023} extend the technique to wrap arbitrary pretrained models without retraining the backbone. The usual motivating intuition is facial recognition: human visual recognition appears to adopt such canonical poses for familiar objects, faces for instance being mentally rotated upright before recognition. In this framework,  the model only needs to learn how to treat the canonical form rather than the full orbit under all symmetry operations.
	
	For crystalline materials, the canonicalization step is comparatively trivial. The symmetry of the underlying point group dictates a small number of well-defined standard bases (specified, for example, in the International Tables for Crystallography), so the canonicalization function need not be learned and instead reduces to a deterministic pre-processing step (implemented in this work via \texttt{spglib} \cite{spglib} and \texttt{pymatgen}), similar to the canonicalization protocol \cite{canonical_tensor}. Once a structure is standardised, both its atomic positions and any tensorial property defined on it can be labelled by point-group irreducible representations directly, allowing the additional specificity exploited in the present work, with components further distinguished within a rotational order by their point-group transformation behaviour.
	
	\subsection*{Related Works}
	Graph neural networks for crystal property prediction range from invariant
	models on scalar features (CGCNN \cite{cgcnn}, SchNet \cite{schnet},
	MEGNet \cite{megnet}, ALIGNN \cite{alignn}, and CGCNN extensions \cite{icgcnn, geocgcnn}), adequate for scalar targets, to $SO(3)$-equivariant
	models that carry features in irreducible representations and couple them with
	Clebsch--Gordan products: tensor field networks \cite{tfn}, the SE(3)-Transformer \cite{se3transformer}, Cormorant \cite{cormorant}, NequIP \cite{nequip},
	MACE \cite{mace}, Allegro \cite{allegro}, EquiformerV2 \cite{equiformerv2}, and
	the \texttt{e3nn} framework \cite{e3nn}. In all of these, equivariance is imposed
	with respect to the \emph{full} rotation group: each rank-$\ell$ feature is a
	single $SO(3)$ irrep with weights shared across all $2\ell+1$ components, so the
	models are agnostic to the crystal's point group and leave any finer site-symmetry
	structure to be learned implicitly from data.

	This machinery has been applied directly to tensorial targets, the closest
	comparison to our work: MatTen \cite{matten} predicts the rank-4 elasticity tensor
	across all seven crystal systems, and related models target dielectric,
	piezoelectric, and Born-charge tensors \cite{anisonet, tensor_pred}. These inherit
	the strict $SO(3)$ equivariance above, emitting the tensor in the arbitrary input
	frame and treating each rank-$\ell$ block as one irrep; they neither canonicalize
	nor exploit the point-group irreducible-representation structure within a
	rotational order that our model is built around.

	An alternative to architectural equivariance is to canonicalize the input, as in
	spatial transformer networks \cite{stn}, learned canonicalization
	\cite{kaba2023, mondal2023}, and frame averaging \cite{frame_averaging}. For
	crystals this is essentially fixed, since the International Tables set a standard
	setting per space group (computed with \texttt{spglib}/\texttt{pymatgen}). We use
	this deterministic standardization not to replace equivariance but to enable a
	finer one: partitioning features and filters by point-group irreducible
	representation in addition to rotational order, recovering the within-$\ell$
	specificity that strictly $SO(3)$-equivariant models cannot express,
	complementary to symmetry-aware graph reductions \cite{congn, wyckoff}.
	
	\subsection*{Structure of this Work}
	The structure of this work is as follows:
	we first provide background on subgroup chains, Neumann's principle, equivariant networks, and point-group harmonics; and then detail the graph construction, standard bases, and the model variants compared in this work. Following this, we present a synthetic demonstration of sub-irrep specificity of symmetry-aware models followed by the parameter-matched four-model comparison on elastic and dielectric tensor prediction from the Materials Project. Finally, we discuss where, why, and when point-group awareness pays in these predictive tasks.
	
	\section*{Background}\label{background}
	\subsection*{Subgroup Chains}
	Symmetry groups are often composed of subgroups, which themselves may have non-trivial subgroups. These relationships may be described succinctly by subgroup chains, allowing for the simultaneous description of some functions transformation properties under this set of groups simultaneously.
	
	The conventional spherical harmonics $Y_{\ell}^m$ are generally labeled according to the subgroup chain $SO(3)\supset SO(2)$, where the rotational order $\ell$ is inherited from the transformation properties under $SO(3)$, and the azimuthal order $m$ is inherited from $SO(2)$ transformations aligned to a conventionally chosen x,y plane.
	
	In symmetric materials with some point group symmetry $G$ however, more apt subgroup chain descriptions are provided by the subgroup chain of this point group. For example, for a material with $O_h$ symmetry, an arbitrary function may be described by it's transformation properties according to the octahedral subgroup chain $
	O_h\supset D_{4h}\supset D_{2h} \supset C_{2h} \supset C_{i}$.
	The transformation properties of a function indexed by this subgroup chain are most often communicated by the irreducible representations (IRs) that they transform under, since arbitrary transformation properties may always be decomposed into a direct product of IRs of the group under consideration.
	As symmetry is lowered, the IR of the higher symmetry super-group becomes an IR of the lower symmetry sub-group, a correspondence often tabulated in so-called 'correlation tables'.

	
	\subsection*{Neumann's Principle}
	Tensorial material properties admit a natural decomposition into irreducible spherical-harmonic blocks of definite rotational rank $\ell$. The rank-2 symmetric dielectric tensor splits into an $\ell = 0$ (isotropic trace) and an $\ell = 2$ (traceless deviatoric) component; the rank-3 piezoelectric tensor splits into $\ell = 1,2$ and $3$ components; the rank-4 elastic tensor splits into $\ell = 0, 2$ and $4$ components. Each block occupies its own $(2\ell + 1)$-dimensional subspace and rotates independently under $SO(3)$, so equivariant networks can process per-$\ell$ channels with separate learnable weights.
	
	Neumann's principle then states that the tensorial properties of a symmetric material must additionally inherit the symmetry of the underlying structure. In a properly labelled IR basis, this means that only independent 1D subspaces indexed by the trivial IR ($A_1$) may be non-zero, since these correspond to the components left unchanged by every symmetry operation of the structure.

	\subsection*{Equivariant Graph Neural Networks for Materials}\label{eq_gnn_bg}
	A natural representation of crystalline materials for machine learning models are graphs: where atomic sites form the nodes and pairs of nearby atoms (those within a cutoff radius, or selected by a local-environment heuristic such as CrystalNN \cite{pymatgen}) form the edges. A graph neural network then operates by iteratively updating each node's feature vector with a learnable function of messages passed from its neighbours along these edges, an architectural primitive known as message passing \cite{mpnn}. The first applications of this idea to crystal property prediction \cite{schnet, cgcnn} treat node and edge features as scalar invariants, an adequate choice for scalar targets such as formation energy and band gap.
	
	Tensorial or directional targets, however, require a stronger property: a rotated input structure must produce a correspondingly rotated prediction. A function $f$ is \textit{equivariant} with respect to a group $G$ acting on its inputs and outputs when, for every $g \in G$,
	\begin{equation}
		f\bigl(\rho_{\mathrm{in}}(g)\,x\bigr) \;=\; \rho_{\mathrm{out}}(g)\,f(x),
	\end{equation}
	where $\rho_{\mathrm{in}}$ and $\rho_{\mathrm{out}}$ are the appropriate representations of $g$ on the input and output spaces. \textit{Invariance} is the special case in which $\rho_{\mathrm{out}}$ is trivial. For materials with no preferred orientation, $G = SO(3)$ is the natural choice, and an $SO(3)$-equivariant network can predict scalar or tensorial properties whose Cartesian components rotate consistently with the structure.
	
	Architectures that enforce $SO(3)$-equivariance through the message-passing layers \cite{tfn, e3nn, se3transformer, o3transformer1} now constitute the standard toolset for predicting orientation-dependent material properties. Reported applications span interatomic potentials \cite{nequip, m3gnet}, full elastic and dielectric tensors \cite{tensor_pred}, and optoelectronic spectra \cite{gnnopt}, with the equivariance guarantee allowing models to generalise to arbitrary rotations of test inputs without data augmentation. The present work extends this line by further partitioning the equivariant operations according to the point group of the underlying structure, exploiting the sub-irrep granularity demonstrated in the synthetic task below.

	\subsection*{Point-Group Harmonics}\label{pgh}
	Within each rank-$\ell$ subspace, the standard $(2\ell+1)$-dimensional basis of real spherical harmonics $\{Y^{\ell}_{m}\}$ (or rank-$\ell$ irreducible pieces of Cartesian tensors of order $\ell$) transforms as a single irreducible representation of $SO(3)$. When restricted to a crystallographic point group $G \subset O(3)$, however, this representation generally reduces into a direct sum of $G$'s irreducible representations. The corresponding basis change is effected by a unitary matrix $U^{(\ell)}_{G}$ that maps the standard basis to a \textit{symmetry-adapted} basis of \textit{point-group harmonics} (PGH) $H^{\ell}_{\alpha i}$, indexed simultaneously by rotational rank $\ell$, point-group irrep $\alpha$, and component index $i$ within that irrep:
	\begin{equation}\label{eq:pgh}
		H^{\ell}_{\alpha i} \;=\; \sum_{m=-\ell}^{\ell}\, \bigl[U^{(\ell)}_{G}\bigr]^{m}_{\alpha i}\, Y^{\ell}_{m}.
	\end{equation}
	For the trivial group $G = \{e\}$ (or equivalently when point-group symmetry is ignored), $U^{(\ell)}_{G}$ is the identity and the PGH basis coincides with the standard one. For any non-trivial $G$, the matrix $U^{(\ell)}_{G}$ \textit{block-diagonalises} every Wigner rotation $D^{(\ell)}(g)$ for $g \in G$: in the PGH basis, components belonging to distinct irreps $\alpha$ are never mixed by symmetry operations of $G$, only with other components of the same irrep. Explicit constructions of $U^{(\ell)}_{G}$ for the crystallographic point groups are tabulated in classical references \cite{bradley_cracknell} and made available algorithmically via the \texttt{MultiPie} library \cite{multipie}, which provides the matrices used throughout this work.
	
	In the standard $Y^{\ell}_{m}$ basis, an $SO(3)$-equivariant operation cannot assign distinct learnable weights to different $m$-components within the same $\ell$ block without breaking equivariance, because the index $m$ carries no $SO(3)$-invariant meaning. In the PGH basis, the IR labels $\alpha$ \textit{are} invariants of the $G$-action by construction, so any operation that respects $G$-equivariance is free to assign distinct weights to components carrying different $\alpha$. The rotational order $\ell$ thereby ceases to be the finest available label for partitioning equivariant weights and features; the pair $(\ell, \alpha)$ takes its place. This is precisely the additional architectural specificity that the present work exploits, and Neumann's principle restated in this basis becomes the statement that the only components of a symmetric crystal property tensor permitted to be non-zero are those with $\alpha = A_1$.

	\section*{Methods}\label{methods}
	\subsection*{Wyckoff Graphs}
	The symmetry of a material often makes atomistic descriptions of it's unit cell structure redundant, since many sites in these symmetric arrangements are essentially equivalent in both chemical and structural environment. The sites equivalent up to the elements of the symmetry group of the material are said to be in the same orbit, and are hence said to occupy the same \textit{Wyckoff position} \cite{wyckoff}. As such, a minimal representation of the occupied sites of a material may be described in terms of these Wyckoff positions, where a multitude of sites with the same interactions (both geometrically and chemically) may be compactly described as one node in a Wyckoff graph. This compact description helps reduce the number of messages passed through equivariant convolution. This is similar to the asymmetric unit cell approach taken in CoNGN \cite{congn}, but with a more direct interpretation in terms of the symmetries of the sites left in the graph. 
	
	In this work, wyckoff graphs were used as model inputs where the graph edges were determined by the CrysNN algorithm as implemented in pymatgen \cite{pymatgen}, which uses a scaled Voronoi tessellation solid angle algorithm (here the default values are used). This algorithm is more selective than the more common crystal graph techniques that only search within some cutoff radius (often large, around 8 Angstrom), both to reduce overfitting and to encourage the formation of edges that most directly reflect the symmetry of the site. These edges' features consisted of spherical harmonics of orders $0\leq\ell\leq \ell_{max}$ evaluated on each edge's corresponding unit vector direction, with the distance between atoms being included as $\ell=0$ features. Initial node features then encoded the following atomic information: atomic number\footnote{\label{onehot}These properties were one-hot encoded.}, group\footref{onehot}, period\footref{onehot}, block\footref{onehot}, electronegativity, atomic radius, number of valence electrons, electron affinity, ionization energy, and atomic mass; following the method adopted in \cite{cgcnn}.

	\subsection*{Standard Bases}
	To accurately label the subspaces of a given material by the IRs they transform as, it's important to first align the symmetry axes with their Cartesian coordinates in some standard fashion. The ITA specifies default axes in their handbook, and this standardization routine is implemented and available in the pymatgen package (or spglib). Adopting a set of standard axes allows for a computationally feasible assignment of IRs to arbitrary subspaces of the material's space and functions on it. A basic check is to confirm that for symmetric arrangements with one independent axis, that this independent axis coincides with that independent component in it's corresponding tensor. Another check is that the symmetry operations that leave the structure unchanged also leave it's corresponding tensor unchanged. 
	
	\begin{figure}[H]
		
		\begin{center}
			\includegraphics[scale=0.33]{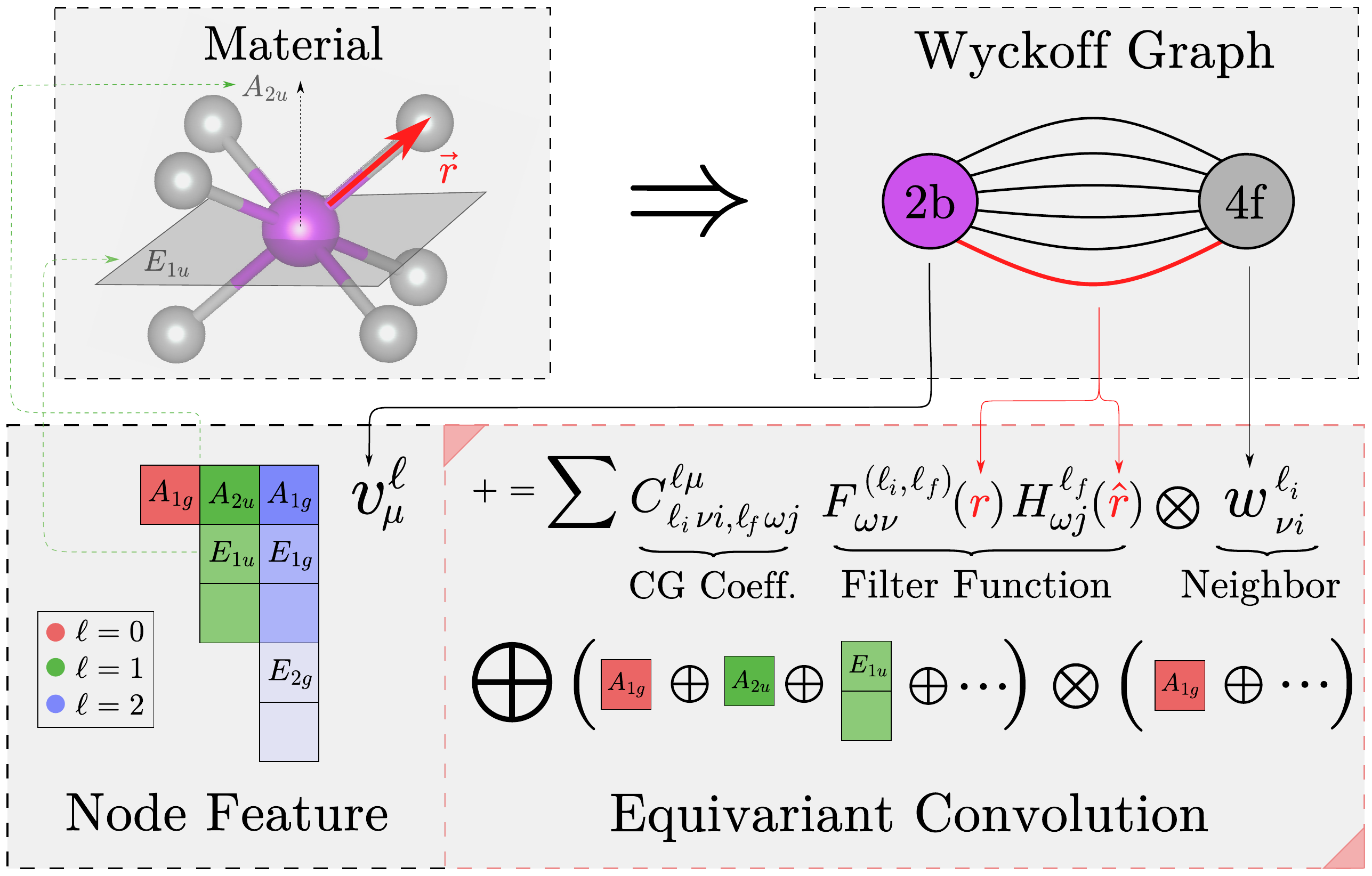}
		\end{center}
		\caption{Symmetry-labelled subspaces of MoS$_2$ and it's corresponding Wyckoff graph. In the present work, symmetric sites are mapped onto one node representing their Wyckoff position in a reduced Wyckoff graph. The central molybdenum sites of the material molybdenum disulfide have $D_{6h}$ symmetry, which may be used to partition the full 3D space into a 2D symmetry plane labelled by the IR $E$, and a trivial axis labelled by the IR $A_1$. Here, $\mathcal{F}$ is an equivariant filter function composed of the product of a point group harmonic $H$ and a learnable radial expansion function $F(r)$ that takes only interatomic distance as argument (as in \cite{tfn}). The feature vector $v$ is updated by it's neighbor via a Clebsch-Gordan decomposition by the coefficients $C$ of the tensor product between this filter function and it's neighboring node features $w$. The indices $\ell, \ell_i,\ell_f$ are rotational orders, $\mu,\omega,\nu$ then point group IR indices, and $i,j,k$ dimensions (if the point group IR isn't one-dimensional) of the origin node feature, neighboring node feature, and the filter function, respectively.}
	\end{figure}

	\subsection*{Invariant Scalar Baseline}\label{a1_baseline}
	Standardizing a structure to its point-group setting has a convenient consequence for the prediction target. By Neumann's principle in the symmetry-adapted basis, only the trivial-IR ($\alpha = A_1$) entries of each rank-$\ell$ block of a symmetric property tensor may be non-zero, and in the canonical frame these entries are fixed scalars. Predicting the tensor therefore reduces to regressing a small, target-specific set of frame-fixed invariants, for which an invariant network suffices: the canonicalization has already absorbed the orientational degrees of freedom that would otherwise demand equivariance. The canonicalization viewpoint is in this sense applied to the output: the model emits the few scalars the tensor is permitted to carry in its standard setting, rather than a  full set of rotating tensor components.
	
	We realize this as a CGCNN-style \cite{cgcnn} invariant baseline. It ingests the same standardized graphs as the equivariant model, using identical CrystalNN edges and atomic node features, but propagates only rotationally-invariant ($\ell = 0$) features, with each edge carrying the interatomic distance alone; invariant message passing, pooling, and an MLP readout emit the $A_1$ coefficients directly. For a like-for-like comparison in Cartesian space, these coefficients are zero-padded into the full $(2\ell+1)$ point-group-harmonic vector, and mapped back to real spherical harmonics by $[U^{(\ell)}_G]^{\top}$, so that both models are scored under one metric. Targets are normalized per rank-$\ell$ block and trained under the same 5-fold cross-validation protocol.
	
	This baseline cleanly isolates the contribution of equivariance. Both models receive the same graphs, the same symmetry information (the point group, its $A_1$ decomposition, and the standardized frame), and predict the same symmetry-allowed scalars; they differ only in whether tensorial ($\ell > 0$) features are constructed and coupled internally through Clebsch--Gordan decomposed products. The gap between the equivariant model and this baseline therefore measures the marginal value of equivariant message passing for the task; that is, whether building and mixing directional intermediate features improves even the invariant outputs, beyond what canonicalization together with an invariant readout already achieve.
		\begin{figure}[H]
		
		\begin{center}
			\includegraphics[scale=0.48]{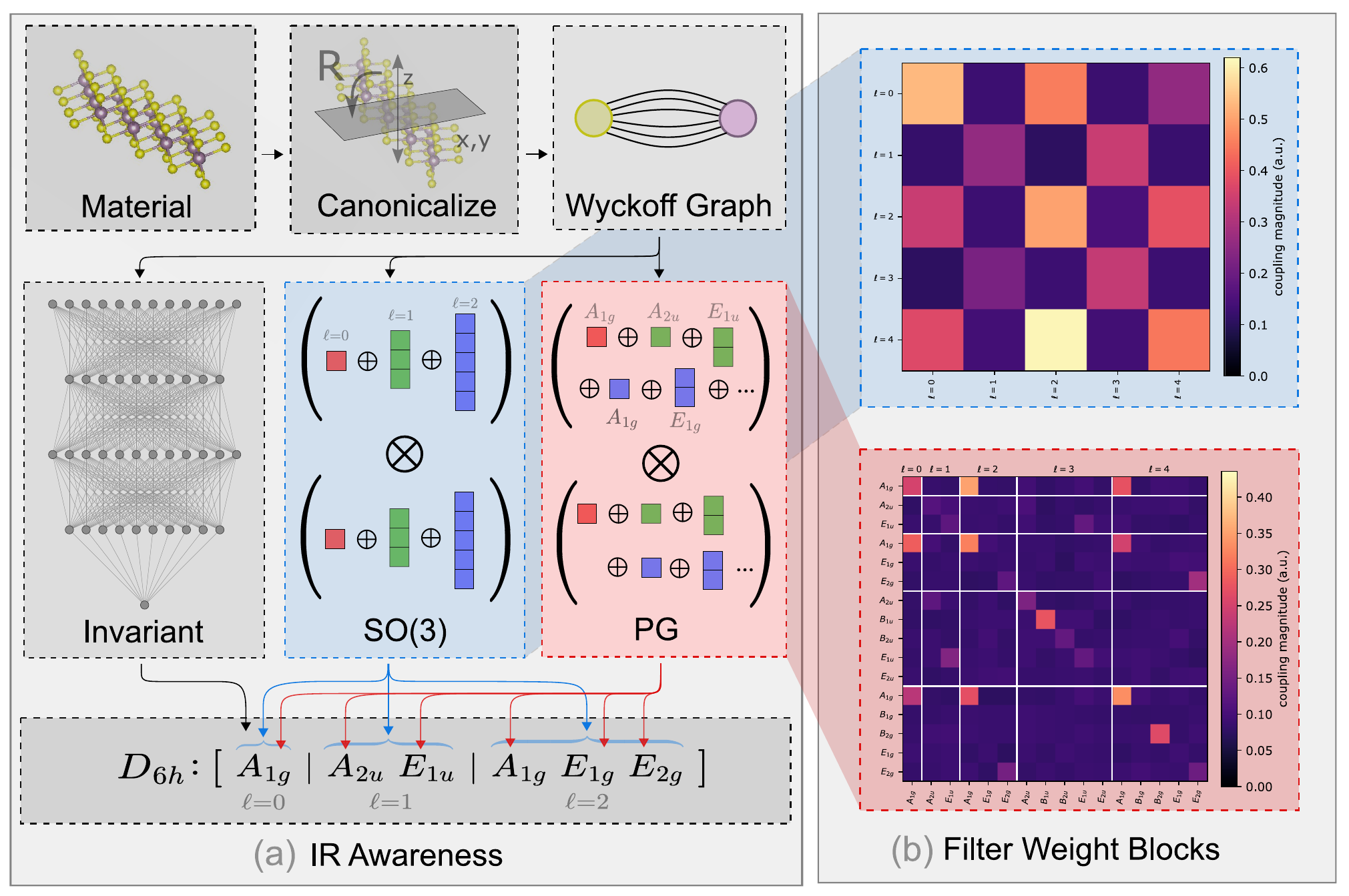}
		\end{center}
		\caption{(a) Workflow for the results presented in this paper. Materials' axis alignment with a standard canonical frame for the materials given point group is performed so that IR labels are appropriately applied. These then are fed into either a standard invariant model (as in \cite{canonical_tensor}), a full $SO(3)$ equivariant model, and a specific, point group (PG) symmetry aware model. These are then used to predict the same sets of target tensors and results are compared. Note that the point group aware model may attribute weights to specific subspace within rotational orders $\ell$ that full rotationally equivariant models cannot, only being able to specify up to $\ell$. (b) Heatmaps showing the Root Mean Squared (RMS) of the weights of each filter block for the last convolutional layer (the second layer of 2) for the architecture used in this work. Note that L2 regularization with $\lambda=$1e-4 was used throughout training of these models to minimize the appearance of spurious weights through backpropagation.}
	\end{figure}

	\subsection*{Symmetry-aware Convolution}
	Since all materials, and their associated properties, are guaranteed at least an equivariant relationship under the rotation group, traditional equivariant convolutional filters are typically indexed by the rotational order $\ell$ inherited from $SO(3)$. These $2\ell +1$ dimensional subspaces are invariant under rotations but grow in size as $\ell$ is increased, so that increasingly large sets of components in these subspaces are indistinguishable to the filter functions and weights of the network.  However, with a point group IR labelled coordinate axis, these subspaces indexed by $\ell$ may be further partitioned into these point group IR labelled subspaces, allowing for more complex relationships to be learned between these IR subspaces, with distinct weights being associated with each as opposed to the more general $\ell$ indexing.
	
\begin{figure}[H]
		\begin{center}
			\includegraphics[scale=0.37]{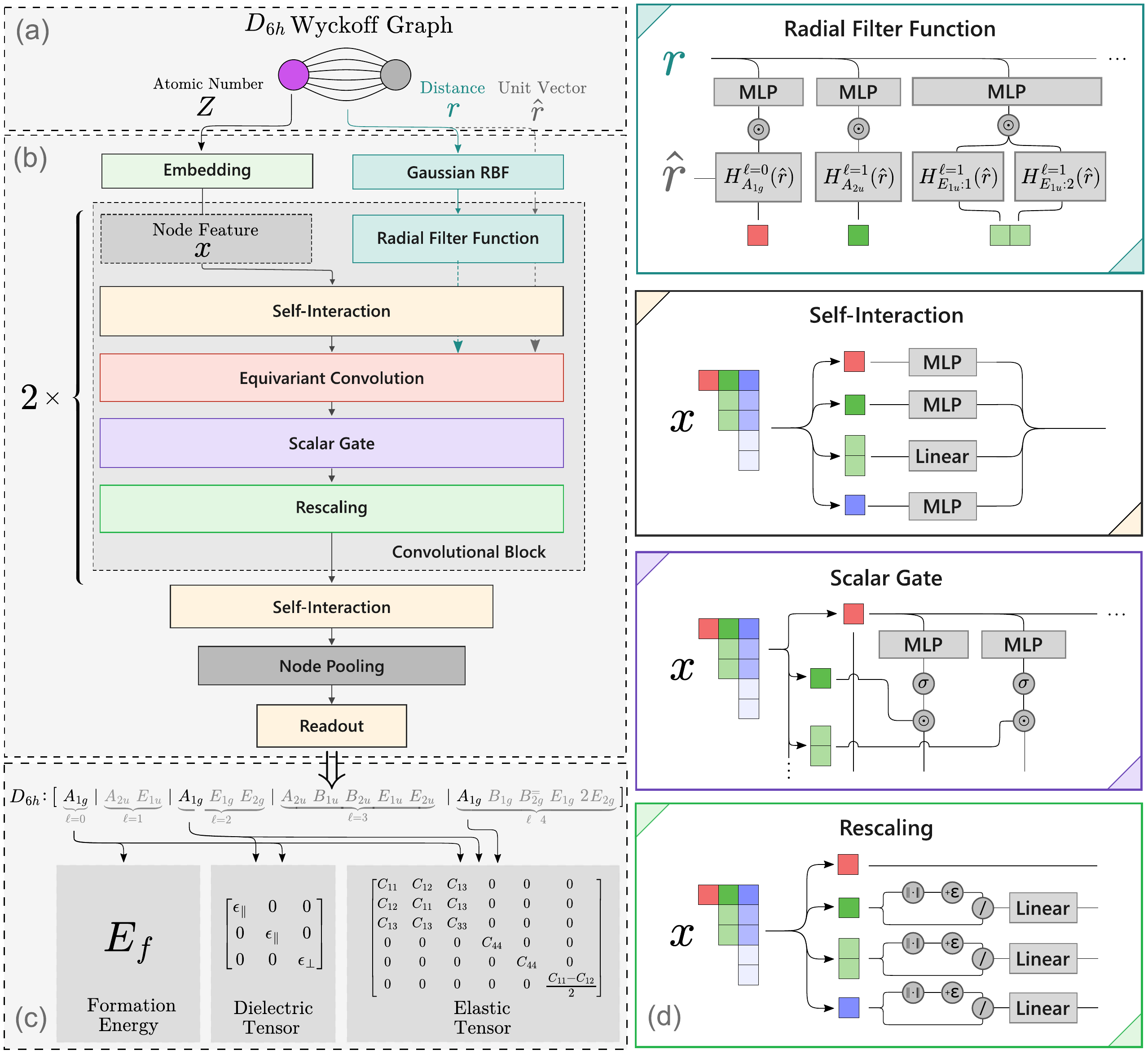}
		\end{center}
		\caption{\textbf{Point Group Equivariant Model Architecture.} Schematic depicting architecture of equivariant models depicted in this work. Atomic embeddings follow that used in CGCNN \cite{cgcnn} for the scalar block, while edge embeddings are a gaussian expansion of distance with dimension equal to block size. For each convolutional block, a radial filter function is evaluated for the edge feature, self interaction is applied to the incoming node features, the decomposed tensor product is then summed with these node features, after which scalar gating between the scalar ($\ell=0$) block and each higher-order block, after which every block is independently normalized and scaled by learned weights. This is then projected down to an output dimension width, node features are pooled over the graph, and a final readout layer is used to predict the target value.}\label{model_arch}
	\end{figure}

	In this case, the feature vectors $(v_{c}^{L})^{\ell\alpha}$ initially representing atomic sites and further associated with nodes throughout the network are labelled not only by their layer $L$, channel $c$, and rotational order $\ell$, but also the point group IR $\alpha$ that they transform under. Note that as long as all point group IRs are maintained, this forms a complete set under which arbitrary features may be decomposed or projected onto.
	The set of equivariant functions already used in equivariant graph neural networks may then be readily adapted to this new basis, namely: equivariant convolution \cite{tfn}, self-interaction \cite{tfn}, scalar gating \cite{equiformerv2}, block and scalar non-linearities \cite{tfn}, and normalization and scaling layers \cite{equiformerv2}. These functions are diagrammatically displayed in \cref{model_arch}.	
	In particular, the convolutional filter $F^{L}_c(r)^{\beta\alpha }_{j}$ of layer $L$, channel $c$, and rotational order $\ell$ may be further indexed by the point group IR label $\beta$ so that message passing then takes the form:
	\begin{equation}
		\big(v^{L+1}_{nc}\big) ^{\gamma}_{n}=\big(v^{L}_{nc}\big)^{\gamma}_n+\sum_{b\in \mathcal{N}(n)}\sum_{\alpha i, \beta j}U_{\alpha i \beta j}^{\gamma n}\big(F^{L}_c(r_{nb})\big)^{\beta\alpha }_{j}\big(v_{bc}^L \big)^{\alpha}_{i}
	\end{equation}
	In this manner, the rotational and point group transformation properties of the feature vectors are both respected and complex interactions between them may be learned.

	\subsection*{$A_1$-Restricted Point-Group Model}\label{a1g_model}
	The learned filter couplings of trained PG-partitioned models motivate a further variant. Inspecting the inter-IR radial couplings of the trained elastic models, the largest magnitudes concentrate on paths into and among the trivial-IR blocks while couplings among non-trivial IRs largely decay toward zero under the $L_2$ penalty, the surviving minority forming a sparse, irrep-matched pattern consisting of precisely those paths whose Clebsch--Gordan products feed back into $A_{1}$ content. Since Neumann's principle already restricts the targets to trivial-IR components, the trained networks thus appear to organize their capacity around exactly the blocks that carry the targets. This motivates a direct test of whether the non-trivial blocks contribute at all.
	
	This $A_1$-restricted model is the PG-partitioned architecture above where the only non-zero learnable weights are confined to the trivial-IR blocks: self-interactions, gates, and readout act only on $A_1$-labelled blocks, and a radial filter is retained only for those pairs in which both members carry the trivial IR. Every feature is then a point-group invariant, a learned coefficient attached to a fixed $A_1$-harmonic pattern, and the model remains (trivially) fully equivariant. It nonetheless retains two capabilities that the neighboring baselines lack. Unlike the invariant scalar baseline, which sees only interatomic distances, it ingests the $A_1$ projections of the edge harmonics at every rotational order, and so remains sensitive to the anisotropy of each local environment through invariant angular combinations. Unlike the $SO(3)$ partition, it carries a separate block, and hence separate weights, normalization, and readout, for each repeated $A_1$ copy within a rotational order: an $SO(3)$-equivariant readout maps an $\ell$-block through a single equivariant linear map and so cannot treat the copies independently, an instance of the within-$\ell$ specificity of the symmetry-aware models.
	Its performance relative to the $SO(3)$ model therefore measures the value of isolating the invariant channel within each rotational order at no additional capacity, while its gap to the full PG model measures the marginal contribution of the non-trivial blocks.

	\section*{Results}\label{results}

	\subsection*{Sub-Irrep Specificity}\label{granularity}
	A direct consequence of the point-group subgroup-chain labelling is that components within the same rotational order $\ell$ are further distinguishable by the point-group IR $\alpha$ that they transform under. Two vector-valued features in a $D_{4h}$ material, for instance, both transform as $\ell = 1$ under $SO(3)$, but decompose further into $A_{2u}$ (the component along the principal $C_4$ axis) and $E_{u}$ (the components within the perpendicular plane) under $D_{4h}$. These are distinct IRs: every point-group operation that fixes the structure leaves the $A_{2u}$ component invariant up to sign while rotating the two $E_{u}$ components into one another, so the two subspaces cannot be mapped onto one another by any equivariant operation.
	
	A standard $SO(3)$-equivariant network is agnostic to this distinction, it sees both vectors as ``the $\ell = 1$ subspace'' and assigns them the same learnable weights. This is acceptable for tasks where the distinction does not matter, but for any task in which the alignment of a directional feature with respect to crystallographic axes is itself the predictive signal, an $SO(3)$-partitioned network of the form used here, whose weights act on the $\ell=1$ block as an indivisible whole and whose readout is invariant, carries no channel in which that alignment can be registered.
	
	To make this concrete, we construct a small synthetic classification task: $300$ tetragonal cells with $D_{4h}$ point-group symmetry are populated with vector-valued node features that are either uniformly along the principal axis ($A_{2u}$) or uniformly in the perpendicular plane ($E_u$), and the target is the binary label of which case applies. The result is shown in \cref{synthetic_table}.

	\begin{figure}[ht]
		\centering
		\includegraphics[scale=0.1]{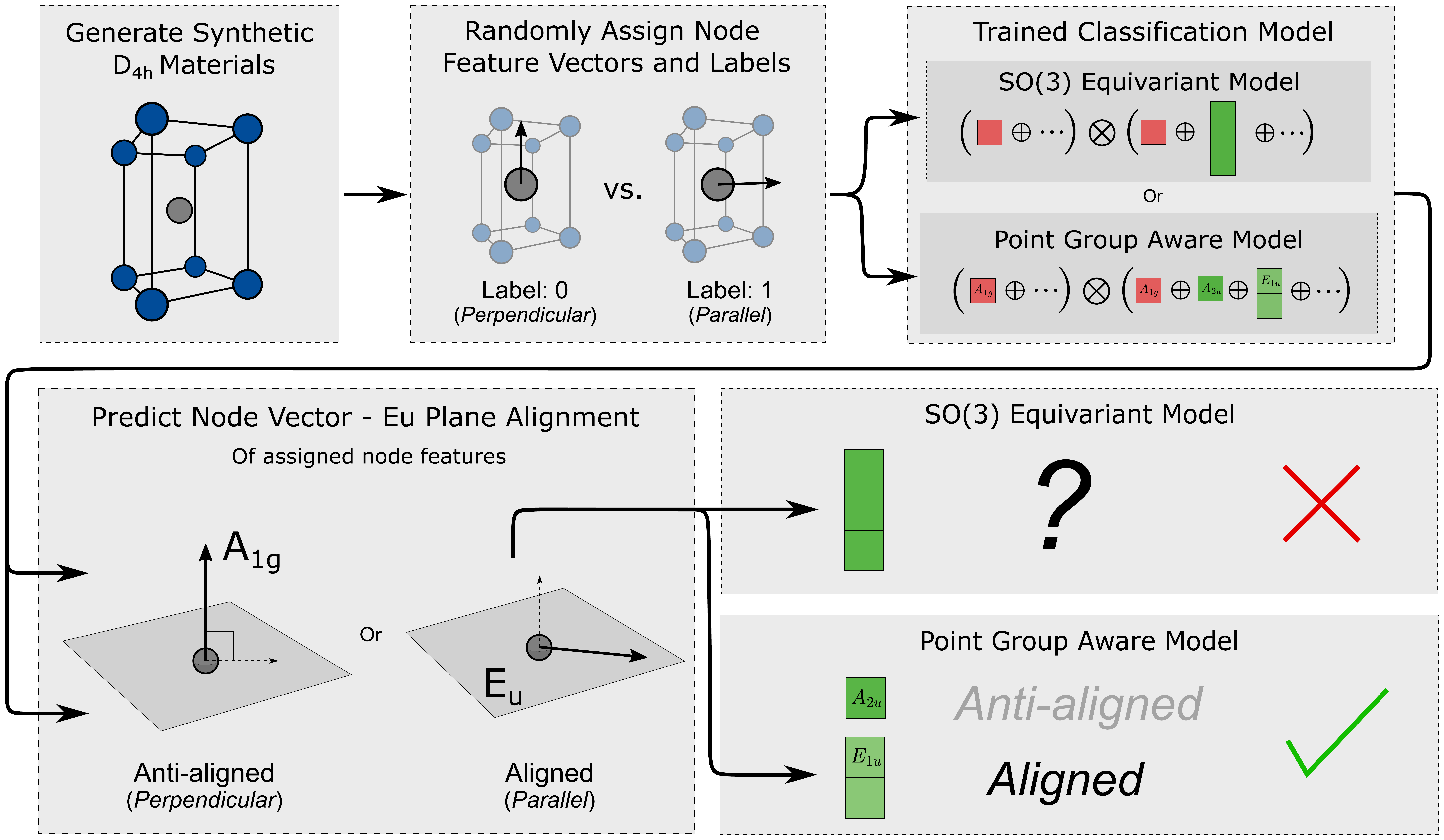}
		\caption{Schematic of the synthetic $D_{4h}$ vector-orientation task. Vector node features are either uniformly along the $A_{2u}$ principal axis or uniformly in the $E_u$ symmetry plane. The $SO(3)$-equivariant network, weighting the $\ell=1$ block as a whole, has no channel that separates the two cases; the PG-partitioned network distinguishes them by routing through separate $A_{2u}$ and $E_u$ weight paths.}\label{synthetic_fig}
	\end{figure}
	
		\begin{table}[ht]
		\centering
		\begin{tabular}{l|l|c|c}
			\toprule
			&	Model & Loss (BCE) & Accuracy \\
			\midrule
			$D_{4h}$ & PG    &  0.000  & 100\%\\
			& SO(3) &	0.584  & 52.5\%\\
			\bottomrule
		\end{tabular}
		\caption{\textbf{$D_{4h}$ Synthetic Training Task.} Classification test results for models with node vectors assigned either parallel to the symmetry plane or perpendicular to it (parallel to the principle axis) where BCE abbreviates binary cross entropy. The alignment of these node vectors is then the class of the synthetic material learned by the model. The test set was 50 of 300 generated $D_{4h}$ materials with 52.5\% of the test set having in-plane node vectors.}\label{synthetic_table}
	\end{table}

	The $SO(3)$ network performs at chance: sharing one weight across the whole $\ell=1$ block and reading out through invariants, it has no channel in which the in-plane and out-of-plane inputs are distinguished. The PG-partitioned network, with separate weights for the $A_{2u}$ and $E_u$ subspaces, solves the task trivially. This architectural distinction, specificity within a rotational order $\ell$, granted by point-group IR labelling, is the foundation of the present work. The remainder of the paper applies the same mechanism to tensor-valued targets on real materials, where the gains are smaller in magnitude but consistent with the same underlying principle.

	\subsection*{Tensor Prediction}
	All data used in this work is drawn from the \textit{Materials Project} database \cite{matproj}. As of database release 2026.04.13, it contains $154{,}377$ non-deprecated materials, nearly all with computed scalar band gap and formation energy values; of these, $11{,}027$ have a non-deprecated computed elastic tensor \cite{elastic-matproj}, $7{,}332$ a computed dielectric tensor \cite{di-matproj}, and $3{,}322$ a computed piezoelectric tensor \cite{piezo-matproj}. Each dataset is split $76\%$ for training and $4\%$ for validation, with the remaining $20\%$ held out for testing in a 5-fold cross-validation scheme: for each fold the model is trained on the training split, and the checkpoint with the best validation performance is used to predict the held-out test set, giving five independent test evaluations.

	In the tensor-prediction comparisons that follow, the three symmetry-aware models are trained side by side on identical folds at a common block width, with capacity equalized by construction rather than by width tuning. The $SO(3)$-partitioned reference sets a parameter budget within every rotational order $\ell$, and each PG variant is trained under fixed binary weight masks, reapplied after every optimizer step, that reduce its active parameter count to this budget: self-interaction weights are matched order by order, and radial filter weights per pair of node and edge orders. The full PG model spreads each order's budget across all of its irrep blocks, while the $A_1$-only variant concentrates the same budget entirely in the trivial-irrep blocks, zeroing every radial filter that touches a non-trivial block; rotational orders that carry no trivial copy forfeit their share outright, so this variant trains strictly leaner than the reference, as recorded in the parameter columns of the tables. This masking removes a confound of comparisons at equal width, where the PG partition necessarily carries more parameters than the $SO(3)$ partition because its parameter count scales with the number of $(\ell, \alpha)$ blocks rather than the number of orders $\ell$; under the matched protocol, any remaining performance difference reflects the partition itself, that is, its irrep resolution, per-block normalization, and per-copy readouts, rather than raw capacity. The invariant scalar baseline stands outside the protocol: it is a plain invariant network that regresses the symmetry-allowed scalars directly in the canonical frame, with no equivariant machinery, and is reported at its native, smaller size.

	\subsubsection*{Elastic Tensor Prediction}
	
	\begin{table}[ht]
		\centering
		\begin{tabular}{l|l|c|ccccc|c}
			\toprule
			\multicolumn{9}{c}{\textbf{Elastic Tensor  Harmonic Component MAE (GPa)}}\\
			\midrule
			Point & Model & Params &\multicolumn{3}{c}{\textit{Symmetric}} & \multicolumn{2}{c}{\textit{Mixed}} & \\
			Group & Partition & ($\times 10^3$) &$\ell=0$ & $\ell=2$ & $\ell=4$ & $\ell=0$ & $\ell=2$ & Total \\
			\midrule
			$O_{h}$ (\textit{3,663})  & PG & 276.9 & 37.6 & 0 & 54.2 & 21.0 & 0 & 22.6 \textit{($\pm$1.1)} \\
			& PG ($A_1$-only) & 133.5 & 37.1 & 0 & 51.8 & 21.5 & 0 & 22.1 \textit{($\pm$1.0)} \\
			& SO(3) & 287.0 & 38.7 & 0 & 52.5 & 21.5 & 0 & 22.5 \textit{($\pm$1.0)} \\
			& Scalar & 84.9 & 39.6 & 0 & 61.8 & 23.4 & 0 & 25.0 \textit{($\pm$1.2)} \\
			\midrule
			$D_{6h}$ (\textit{1,033})   & PG & 291.9 & 52.3 & 46.5 & 51.5 & 30.3 & 19.6 & 40.0 \textit{($\pm$3.1)} \\
			& PG ($A_1$-only) & 212.6 & 51.4 & 46.5 & 51.3 & 31.2 & 19.2 & 39.9 \textit{($\pm$3.0)} \\
			& SO(3) & 287.0 & 52.8 & 48.2 & 52.6 & 30.0 & 20.3 & 40.8 \textit{($\pm$2.5)} \\
			& Scalar & 85.0 & 59.4 & 59.5 & 65.9 & 42.3 & 29.1 & 51.2 \textit{($\pm$4.0)} \\
			\midrule
			$D_{4h}$ (\textit{1,701}) & PG & 294.0 & 39.3 & 29.0 & 45.1 & 31.6 & 18.0 & 32.6 \textit{($\pm$1.8)} \\
			& PG ($A_1$-only) & 214.0 & 38.7 & 28.5 & 42.8 & 31.7 & 17.8 & 31.9 \textit{($\pm$1.7)} \\
			& SO(3) & 287.0 & 37.8 & 29.6 & 42.6 & 31.0 & 19.3 & 32.0 \textit{($\pm$2.1)} \\
			& Scalar & 85.1 & 43.1 & 35.8 & 57.9 & 37.9 & 26.8 & 40.3 \textit{($\pm$1.9)} \\
			\midrule
			$D_{2h}$ (\textit{1,254}) & PG & 294.5 & 40.6 & 37.3 & 45.9 & 34.3 & 23.0 & 36.2 \textit{($\pm$1.9)} \\
			& PG ($A_1$-only) & 217.4 & 39.9 & 36.4 & 45.4 & 33.1 & 22.0 & 35.4 \textit{($\pm$1.2)} \\
			& SO(3) & 287.0 & 43.6 & 39.4 & 52.4 & 34.9 & 24.1 & 38.9 \textit{($\pm$1.7)} \\
			& Scalar & 85.3 & 53.0 & 41.4 & 54.3 & 45.9 & 26.1 & 44.1 \textit{($\pm$2.2)} \\
			\bottomrule
		\end{tabular}
		\caption{ \textbf{Equal Parameter Elastic Models.} Harmonic component Mean Absolute Error (MAE) on elastic tensor datasets from Materials Project of four point group symmetries. Results are recorded in $GPa$, with the cross-validation standard deviation of the total MAE italicized in parentheses. The Params column gives each model's active parameter count in thousands: the PG and PG ($A_1$-only) variants are budget-matched to the SO(3) reference per rotational order, with the $A_1$-only model forfeiting the budget of rotational orders that carry no trivial-irrep block, while the scalar baseline is a smaller invariant architecture. Models are masked according to symmetry classes so MAE is only over non-zero components. The MAD baseline row reports the error of predicting the per-component dataset mean (mean absolute deviation), the skill-free reference for each block; its total is the same average of per-block errors used for the model rows. Number of data in each symmetry dataset is given in italics. }
	\end{table}
	
	At matched parameter budgets the three symmetry-aware partitions are closely grouped on every dataset and all sit far below the MAD baseline in total, while the smaller scalar baseline trails them throughout. The $A_1$-only restriction attains the lowest total MAE on all four datasets despite training the fewest active parameters of the three, and the full PG partition separates clearly from SO(3) only for $D_{2h}$.
	
	For cubic $O_h$ the symmetry-aware models are separated by little beyond the fold scatter, with total MAEs of $22.6$ GPa for PG, $22.1$ for the $A_1$-only variant, and $22.5$ for SO(3) against $25.0$ for the scalar baseline and a MAD floor of $82.1$. The symmetric $\ell=0$ block spans $37.1$ to $39.6$ GPa across the four models. The symmetric $\ell=4$ block is the hardest, at $54.2$ GPa for PG, $51.8$ for $A_1$-only, $52.5$ for SO(3), and $61.8$ for the scalar model, against a floor of $122.9$. The mixed $\ell=0$ block sits at $21.0$, $21.5$, $21.5$, and $23.4$ GPa respectively. Both $\ell=2$ blocks are identically zero by cubic symmetry, so the point group exposes no additional within-order structure to exploit; the $A_1$-only model nonetheless posts the best total with $133$ thousand active parameters, less than half the SO(3) budget, since cubic symmetry admits no trivial copy at $\ell=1,2,3$ and the corresponding budget is forfeited.
	
	For $D_{6h}$, the highest-symmetry anisotropic group, the totals are $40.0$ GPa for PG and $39.9$ for $A_1$-only against $40.8$ for SO(3), $51.2$ for the scalar baseline, and a MAD floor of $113.6$. The symmetric $\ell=0$ block is $52.3$ GPa for PG, $51.4$ for $A_1$-only, $52.8$ for SO(3), and $59.4$ for the scalar model. The symmetric $\ell=2$ block is $46.5$, $46.5$, $48.2$, and $59.5$ GPa against a floor of $74.7$. The symmetric $\ell=4$ block is $51.5$, $51.3$, $52.6$, and $65.9$ GPa against a floor of $85.2$. The mixed $\ell=0$ block is $30.3$, $31.2$, $30.0$, and $42.3$ GPa, and the mixed $\ell=2$ block is $19.6$, $19.2$, $20.3$, and $29.1$ GPa against a floor of $39.1$. The scalar deficit is widest on this dataset and appears at every block.
	
	For $D_{4h}$, the one dataset where SO(3) edges the full PG partition in total, the ordering is $31.9$ GPa for $A_1$-only, $32.0$ for SO(3), and $32.6$ for PG against $40.3$ for the scalar baseline and a MAD floor of $80.1$. The symmetric $\ell=0$ block is $39.3$, $38.7$, $37.8$, and $43.1$ GPa for the PG, $A_1$-only, SO(3), and scalar models. The symmetric $\ell=2$ block is $29.0$, $28.5$, $29.6$, and $35.8$ GPa against a floor of $48.0$. The symmetric $\ell=4$ block is $45.1$, $42.8$, $42.6$, and $57.9$ GPa against a floor of $76.1$. The mixed $\ell=0$ block is $31.6$, $31.7$, $31.0$, and $37.9$ GPa, and the mixed $\ell=2$ block is $18.0$, $17.8$, $19.3$, and $26.8$ GPa against a floor of $34.9$.
	
	For $D_{2h}$ the PG family separates most clearly from the SO(3) reference, with totals of $36.2$ GPa for PG and $35.4$ for $A_1$-only against $38.9$ for SO(3), $44.1$ for the scalar baseline, and a MAD floor of $86.1$. The symmetric $\ell=0$ block is $40.6$, $39.9$, $43.6$, and $53.0$ GPa for the PG, $A_1$-only, SO(3), and scalar models. The symmetric $\ell=2$ block sits almost at its no-skill floor of $38.4$ GPa: the PG variants extract a small signal at $37.3$ and $36.4$, while SO(3) and the scalar model land above the floor at $39.4$ and $41.4$, no better than predicting the per-component mean. The symmetric $\ell=4$ block shows the same pattern less severely, at $45.9$ and $45.4$ GPa for the PG variants against $52.4$ for SO(3), $54.3$ for the scalar model, and a floor of $56.1$. The mixed $\ell=0$ block is $34.3$, $33.1$, $34.9$, and $45.9$ GPa, and the mixed $\ell=2$ block is again near its floor of $24.3$, at $23.0$, $22.0$, $24.1$, and $26.1$ GPa, with only the PG variants below it.
	
	Three regularities hold across groups. First, the scalar baseline trails every symmetry-aware model on every dataset, confirming that the invariant angular content of the local environments carries predictive information that interatomic distances alone do not. Second, the $A_1$-only restriction matches or improves on the full PG partition throughout, indicating that the predictive content of the point-group partition is concentrated in its trivial-irrep blocks. Third, where the anisotropic signal is faint, as in the $D_{2h}$ $\ell=2$ blocks whose floors sit within a few GPa of the best models, the trivial-block isolation of the PG family still recovers a measurable signal while the SO(3) partition does not improve on the no-skill floor.

	\subsubsection*{Dielectric Tensor Prediction}
	
	\begin{table}[ht]
		\centering
		\begin{tabular}{l|l|c|cc|c}
			\toprule
			\multicolumn{6}{c}{\textbf{Dielectric Tensor  Harmonic Component MAE }}\\
			\midrule
			Point & Model & Params & \multicolumn{2}{c}{\textit{Spherical Component}} & \\
			Group & Partition & ($10^3$) &$\ell=0$ & $\ell=2$ &  Total \\
			\midrule
			$D_{2h}$ (\textit{602})  & PG & 164.5 & 0.479 & 0.376 & 0.428 \textit{($\pm$0.048)} \\
			& PG ($A_1$-only) & 134.5 & 0.501 & 0.369 & 0.435 \textit{($\pm$0.042)} \\
			& SO(3) & 162.2 & 0.451 & 0.406 & 0.428 \textit{($\pm$0.042)} \\
			& Scalar & 84.9 & 0.566 & 0.393 & 0.480 \textit{($\pm$0.057)} \\
			\midrule
			$D_{3d}$ (\textit{330})   & PG & 163.2 & 0.478 & 0.314 & 0.396 \textit{($\pm$0.028)} \\
			& PG ($A_1$-only) & 133.3 & 0.510 & 0.339 & 0.425 \textit{($\pm$0.046)} \\
			& SO(3) & 162.2 & 0.542 & 0.400 & 0.471 \textit{($\pm$0.078)} \\
			& Scalar & 84.8 & 0.552 & 0.353 & 0.452 \textit{($\pm$0.058)} \\
			\midrule
			$C_{2v}$ (\textit{424}) & PG & 165.0 & 0.492 & 0.221 & 0.357 \textit{($\pm$0.048)} \\
			& PG ($A_1$-only) & 164.0 & 0.482 & 0.218 & 0.350 \textit{($\pm$0.042)} \\
			& SO(3) & 162.2 & 0.491 & 0.224 & 0.357 \textit{($\pm$0.045)} \\
			& Scalar & 84.9 & 0.517 & 0.263 & 0.390 \textit{($\pm$0.049)} \\
			\midrule
			$C_{2h}$ (\textit{655}) & PG & 164.5 & 0.542 & 0.252 & 0.397 \textit{($\pm$0.012)} \\
			& PG ($A_1$-only) & 135.2 & 0.541 & 0.253 & 0.397 \textit{($\pm$0.044)} \\
			& SO(3) & 162.2 & 0.531 & 0.261 & 0.396 \textit{($\pm$0.019)} \\
			& Scalar & 84.9 & 0.545 & 0.335 & 0.440 \textit{($\pm$0.017)} \\
			\bottomrule
		\end{tabular}
		\caption{ \textbf{Equal Parameter Dielectric Models.} Harmonic component Mean Absolute Error (MAE) on dielectric tensor datasets from Materials Project of four point group symmetries. Recorded results are unitless, with the cross-validation standard deviation of the total MAE italicized in parentheses. The Params column gives each model's active parameter count in thousands: the PG and PG ($A_1$-only) variants are budget-matched to the SO(3) reference per rotational order, with the $A_1$-only model forfeiting the budget of rotational orders that carry no trivial-irrep block, while the scalar baseline is a smaller invariant architecture. Models are masked according to symmetry classes so MAE is only over non-zero components. The MAD baseline row reports the error of predicting the per-component dataset mean (mean absolute deviation), the skill-free reference for each block; its total is the same average of per-block errors used for the model rows. Number of data in each symmetry dataset is given in italics. }
	\end{table}
	
	
	The dielectric tensor is rank two, splitting into an isotropic $\ell=0$ trace and an anisotropic $\ell=2$ deviatoric block. At matched parameter budgets the ordering differs from the elastic case: the three symmetry-aware partitions are statistically indistinguishable in total on three of the four groups, and a clear separation appears only for $D_{3d}$, where the full PG partition is strongest. Every model recovers the $\ell=0$ trace to roughly a third of its MAD floor, while the $\ell=2$ block carries meaningful learnable signal only for $D_{3d}$ and, weakly, $D_{2h}$.
	
	For $D_{2h}$ the totals are $0.428$ for PG, $0.435$ for $A_1$-only, and $0.428$ for SO(3) against $0.480$ for the scalar baseline and a MAD floor of $0.999$. The $\ell=0$ block is $0.479$, $0.501$, $0.451$, and $0.566$ for the PG, $A_1$-only, SO(3), and scalar models against a floor of $1.573$. The $\ell=2$ block is $0.376$, $0.369$, $0.406$, and $0.393$ against a floor of $0.425$, a weak anisotropic signal of which the PG variants recover the most.
	
	For $D_{3d}$ the full PG partition separates clearly, with a total of $0.396$ against $0.425$ for $A_1$-only, $0.471$ for SO(3), $0.452$ for the scalar baseline, and a MAD floor of $1.115$; this is also the only dataset where even the scalar baseline matches SO(3). The $\ell=0$ block is $0.478$ for PG, $0.510$ for $A_1$-only, $0.542$ for SO(3), and $0.552$ for the scalar model. The $\ell=2$ block carries the strongest learnable anisotropic signal of the set, $0.314$ for PG against $0.339$, $0.400$, and $0.353$ for the $A_1$-only, SO(3), and scalar models, on a floor of $0.547$. The PG lead appears at both components, though the fold scatter here is also the largest of the set.
	
	For $C_{2v}$ the three symmetry-aware models are tied in total at $0.357$, $0.350$, and $0.357$ for PG, $A_1$-only, and SO(3) against $0.390$ for the scalar baseline and a floor of $0.795$. The $\ell=0$ block is $0.492$, $0.482$, $0.491$, and $0.517$ against a floor of $1.359$. The $\ell=2$ block sits essentially at its no-skill floor of $0.231$ for every symmetry-aware model, at $0.221$, $0.218$, and $0.224$, with the scalar model above it at $0.263$.
	
	For $C_{2h}$ the totals are likewise tied at $0.397$, $0.397$, and $0.396$ against $0.440$ for the scalar baseline and a floor of $0.901$. The $\ell=0$ block is $0.542$, $0.541$, $0.531$, and $0.545$ against a floor of $1.545$. The $\ell=2$ block again sits at its floor of $0.258$ for the symmetry-aware models, at $0.252$, $0.253$, and $0.261$, while the scalar model lands well above it at $0.335$.
	
	Taken together, the dielectric results temper the elastic picture. Most of the predictable content of the dielectric tensor is isotropic, and for the monoclinic and orthorhombic groups the deviatoric block offers little signal beyond its floor to any partition, so the partitions tie. Where a genuinely learnable anisotropic block exists, in the trigonal $D_{3d}$ set, the point-group partition exploits it most fully, consistent with the sub-irrep specificity of the symmetry-aware models.

	\section*{Discussion}\label{discuss}
	The parameter-matched four-model comparison dissects point-group awareness into its separable ingredients, and the resulting picture is more differentiated than a blanket advantage. The angular content itself is clearly necessary: the scalar baseline trails every symmetry-aware model on every dataset, and its deficit is not a capacity artifact, since on several anisotropic blocks it lands above the no-skill MAD floor (for instance $0.335$ against a floor of $0.258$ on the $C_{2h}$ dielectric $\ell=2$ block). Among the three symmetry-aware partitions, however, the totals are closely grouped, and the $A_1$-only restriction is never worse than the full PG partition on the elastic sets while matching it on three of the four dielectric sets, despite training the fewest active parameters of the three. The predictive content of the point-group partition is thus concentrated in its trivial-irrep blocks, confirming causally what the learned filter couplings of the full models already suggested.
	
	This concentration is intelligible from Neumann's principle. The masked targets are purely trivial-irrep objects, so a model that ingests the $A_1$ projections of the edge harmonics at every order, and carries a separate block for every repeated $A_1$ copy within an order, already commands the invariant angular combinations from which the targets are built; the non-trivial blocks can contribute only indirectly, through product paths that return to $A_1$ content. The comparison shows that at these dataset sizes such indirect contributions are mostly negligible. What the trivial blocks provide beyond the $SO(3)$ partition is the within-order specificity in its minimal form: separate weights, normalization, and readout heads for each $A_1$ copy, a resolution an $SO(3)$-equivariant readout cannot express.
	
	The comparison equally shows where the fuller partition earns its keep, and the MAD floors locate the pattern. The two datasets with a clear PG-family advantage are precisely those whose anisotropic blocks carry signal in a nontrivial regime. For the $D_{3d}$ dielectric set the $\ell=2$ block holds the strongest learnable anisotropic signal of the study, and there the full PG partition beats not only SO(3) ($0.396$ against $0.471$ in total) but also its own $A_1$-only restriction ($0.425$), the one case where the non-trivial blocks demonstrably add value. For the $D_{2h}$ elastic set the anisotropic signal is instead faint, with $\ell=2$ floors within a few GPa of the best models, and there the PG variants stay at or below the floors while the $SO(3)$ model drifts above them. We read the latter as an interference effect: the $SO(3)$ model's $\ell=2$ output rides on filters shared across all components of the order and on features that also feed the large $\ell=0$ losses, so a faint two-component signal is dragged by gradient traffic it cannot escape, whereas the PG partition isolates the trivial copies in dedicated blocks. 
	
	The cubic $O_h$ result remains the control. Its elastic tensor carries no $\ell=2$ content and the cubic group leaves a single trivial copy at $\ell=4$, so the performance of all three symmetry-aware partitions are statistically indistinguishable there; notably, the $A_1$-only model matches them with less than half the active parameters, since the orders with no trivial copy forfeit their budget. Across the study, then, the point-group partition helps under three coinciding conditions: the target populates $\ell>0$ subspaces, the group resolves those subspaces into multiple irreps or repeated trivial copies, and the dataset actually carries anisotropic signal above its no-skill floor. For much of the available tensor data the last condition is the binding one, and the practical yield of point-group awareness is correspondingly less a large accuracy gain than a leaner model of equal accuracy and a novel direct per-copy prediction of the symmetry-allowed components.

	\section*{Conclusion}
	In this work we introduced point-group symmetry-aware equivariant networks, which refine the conventional rotational-order partitioning of equivariant architectures into the finer labelling afforded by a material's point group. By standardizing each structure to a canonical setting and partitioning features, weights, and filter functions by point-group irreducible representation in addition to rotational order, these models acquire a sub-irrep specificity that purely $SO(3)$-equivariant networks are forbidden by symmetry to express, all while retaining full equivariance. Applied to the prediction of tensorial material properties under a parameter-matched protocol, this granularity proves valuable in a specific sense rather than a broad increase in accuracy: its predictive content concentrates in the trivial-irrep blocks, so that an $A_1$-restricted variant matches or improves on both the full partition and the $SO(3)$ reference with fewer active parameters, while the full partition pays off where the data carry genuinely learnable anisotropic signal. The point-group labels thus buy leaner models of equal accuracy, direct prediction of each symmetry-allowed component, and robustness on weak-signal blocks. Looking ahead, the same construction extends naturally to a broader range of directional and tensorial properties, to lower-symmetry and magnetic point groups, and to the larger, more diverse datasets on which symmetry-aware inductive biases are likely to prove most valuable; more broadly, it suggests that explicitly encoding the discrete symmetries of a system, and not only its continuous ones, is a productive direction for physically grounded machine learning.
	
	Future work may explore greater applicability of the symmetry-adapted weights through weight sharing informed by correlation tables, or larger generated datasets that probe these models beyond the small symmetry-resolved cohorts available today. The deeper limitation, however, lies with the targets themselves: by Neumann's principle a crystal property tensor exercises only the trivial-irrep channel, so the non-invariant output blocks are summarily discarded and tensor prediction cannot exhibit the full potential of the IR-labelled specificity. Targets that genuinely populate non-trivial irreps, such as electronic Hamiltonians, are the natural setting in which the finer partition may yet show its full potential.
	
	\printbibliography

@inproceedings{se3transformer,
	title={{SE(3)}-Transformers: {3D} roto-translation equivariant attention networks},
	author={Fuchs, Fabian B. and Worrall, Daniel E. and Fischer, Volker and Welling, Max},
	booktitle={Advances in Neural Information Processing Systems (NeurIPS)},
	volume={33}, year={2020}
}

@inproceedings{cormorant,
	title={Cormorant: Covariant molecular neural networks},
	author={Anderson, Brandon and Hy, Truong Son and Kondor, Risi},
	booktitle={Advances in Neural Information Processing Systems (NeurIPS)},
	volume={32}, year={2019}
}

@inproceedings{mace,
	title={{MACE}: Higher order equivariant message passing neural networks for fast and accurate force fields},
	author={Batatia, Ilyes and Kov{\'a}cs, D{\'a}vid P{\'e}ter and Simm, Gregor N. C. and Ortner, Christoph and Cs{\'a}nyi, G{\'a}bor},
	booktitle={Advances in Neural Information Processing Systems (NeurIPS)},
	volume={35}, year={2022}
}

@article{allegro,
	title={Learning local equivariant representations for large-scale atomistic dynamics},
	author={Musaelian, Albert and Batzner, Simon and Johansson, Anders and Sun, Lixin and Owen, Cameron J. and Kornbluth, Mordechai and Kozinsky, Boris},
	journal={Nature Communications}, volume={14}, number={1}, pages={579}, year={2023},
	publisher={Nature Publishing Group}
}

@article{matten,
	title={An equivariant graph neural network for the elasticity tensors of all seven crystal systems},
	author={Wen, Mingjian and Horton, Matthew K. and Munro, Jason M. and Huck, Patrick and Persson, Kristin A.},
	journal={Digital Discovery}, volume={3}, number={5}, pages={869--882}, year={2024},
	publisher={Royal Society of Chemistry}
}

@inproceedings{frame_averaging,
	title={Frame averaging for invariant and equivariant network design},
	author={Puny, Omri and Atzmon, Matan and Ben-Hamu, Heli and Misra, Ishan and Grover, Aditya and Smidt, Edward J. and Lipman, Yaron},
	booktitle={International Conference on Learning Representations (ICLR)}, year={2022}
}

@article{anisonet,
	title = {Discovery of Highly Anisotropic Dielectric Crystals with Equivariant Graph Neural Networks},
	author = {Yuchen Lou and Alex M. Ganose},
	year = {2025},
	journal = {Faraday Discussions},
	volume = {256},
	number = {0},
	pages = {255--274},
	doi = {10.1039/D4FD00096J},
	url = {https://pubs.rsc.org/en/content/articlelanding/2025/fd/d4fd00096j},
}

@inproceedings{stn,
	title     = {Spatial Transformer Networks},
	author    = {Jaderberg, Max and Simonyan, Karen and Zisserman, Andrew and Kavukcuoglu, Koray},
	booktitle = {Advances in Neural Information Processing Systems (NeurIPS)},
	volume    = {28},
	year      = {2015}
}

@inproceedings{kaba2023,
	title     = {Equivariance with Learned Canonicalization Functions},
	author    = {Kaba, S{\'{e}}kou-Oumar and Mondal, Arnab Kumar and Zhang, Yan and Bengio, Yoshua and Ravanbakhsh, Siamak},
	booktitle = {Proceedings of the 40th International Conference on Machine Learning (ICML)},
	series    = {Proceedings of Machine Learning Research},
	volume    = {202},
	pages     = {15546--15566},
	year      = {2023}
}

@inproceedings{mondal2023,
	title     = {Equivariant Adaptation of Large Pretrained Models},
	author    = {Mondal, Arnab Kumar and Panigrahi, Siba Smarak and Kaba, S{\'{e}}kou-Oumar and Mudumba, Sai Rajeshwar and Ravanbakhsh, Siamak},
	booktitle = {Advances in Neural Information Processing Systems (NeurIPS)},
	volume    = {36},
	year      = {2023}
}

@article{congn,
  title={Connectivity optimized nested line graph networks for crystal structures},
  author={Ruff, Robin and Reiser, Patrick and St{\"u}hmer, Jan and Friederich, Pascal},
  journal={Digit. Discov.},
  volume={3},
  number={3},
  pages={pp. 594--601},
  year={2024},
  publisher={Royal Society of Chemistry}
}

@article{matproj,
    author = {Jain, Anubhav and Ong, Shyue Ping and Hautier, Geoffroy and Chen, Wei and Richards, William Davidson and Dacek, Stephen and Cholia, Shreyas and Gunter, Dan and Skinner, David and Ceder, Gerbrand and Persson, Kristin A.},
    title = {Commentary: The Materials Project: A materials genome approach to accelerating materials innovation},
    journal = {APL Mater.},
    volume = {1},
    number = {1},
    pages = {article 011002},
    year = {2013},
    month = {07},
    issn = {2166-532X},
    doi = {10.1063/1.4812323},
    url = {https://doi.org/10.1063/1.4812323},
    eprint = {https://pubs.aip.org/aip/apm/article-pdf/doi/10.1063/1.4812323/13163869/011002\_1\_online.pdf},
}

@article{schnet,
 author = {Schütt, K. T. and Sauceda, H. E. and Kindermans, P.-J. and Tkatchenko, A. and Müller, K.-R.},
 title = {SchNet – A deep learning architecture for molecules and materials},
 journal = {J. Chem. Phys.},
 volume = {148},
 number = {24},
 pages = {article 241722},
 year = {2018},
 month = {03},
 issn = {0021-9606},
 doi = {10.1063/1.5019779},
 url = {https://doi.org/10.1063/1.5019779},
 eprint = {https://pubs.aip.org/aip/jcp/article-pdf/doi/10.1063/1.5019779/16655678/241722\_1\_online.pdf},
 }

@article{cgcnn,
  title={Crystal graph convolutional neural networks for an accurate and interpretable prediction of material properties},
  author={Xie, Tian and Grossman, Jeffrey C},
  journal={{PRL}},
  volume={120},
  number={14},
  pages={article 145301},
  year={2018},
  publisher={APS}
}

@article{m3gnet,
  title={A universal graph deep learning interatomic potential for the periodic table},
  author={Chen, Chi and Ong, Shyue Ping},
  journal={Nat. Comput. Sci.},
  volume={2},
  number={11},
  pages={pp.  718--728},
  year={2022},
  publisher={Nature Publishing Group US New York}
}

@article{alignn,
  title={Atomistic Line Graph Neural Network for improved materials property predictions},
  author={Choudhary, Kamal and DeCost, Brian},
  journal={{npj} Comput. Mater.},
  volume={7},
  number={1},
  pages={pp.  1--8},
  year={2021},
  publisher={Nature Publishing Group}
}

@article{nequip,
	title     = {{E(3)}-equivariant graph neural networks for data-efficient and accurate interatomic potentials},
	author    = {Batzner, Simon and Musaelian, Albert and Sun, Lixin and Geiger, Mario and Mailoa, Jonathan P. and Kornbluth, Mordechai and Molinari, Nicola and Smidt, Tess E. and Kozinsky, Boris},
	journal   = {Nature Communications},
	volume    = {13},
	number    = {1},
	pages     = {2453},
	year      = {2022},
	doi       = {10.1038/s41467-022-29939-5}
}

@book{bradley_cracknell,
	title     = {The Mathematical Theory of Symmetry in Solids: Representation Theory for Point Groups and Space Groups},
	author    = {Bradley, Christopher J. and Cracknell, Arthur P.},
	publisher = {Clarendon Press, Oxford},
	year      = {1972},
	isbn      = {9780198519201}
}

@article{multipie,
	title   = {Multipole expansion for magnetic structures: A generation scheme for a symmetry-adapted orthonormal basis set in the crystallographic point group},
	author  = {Suzuki, M.-T. and Nomoto, T. and Arita, R. and Yanagi, Y. and Hayami, S. and Kusunose, H.},
	journal = {Physical Review B},
	volume  = {99},
	number  = {17},
	pages   = {174407},
	year    = {2019},
	doi     = {10.1103/PhysRevB.99.174407}
}

@article{di-matproj,
	author={Petousis, Ioannis
	and Mrdjenovich, David
	and Ballouz, Eric
	and Liu, Miao
	and Winston, Donald
	and Chen, Wei
	and Graf, Tanja
	and Schladt, Thomas D.
	and Persson, Kristin A.
	and Prinz, Fritz B.},
	title={High-throughput screening of inorganic compounds for the discovery of novel dielectric and optical materials},
	journal={Scientific Data},
	year={2017},
	month={Jan},
	day={31},
	volume={4},
	number={1},
	pages={160134},
	issn={2052-4463},
	doi={10.1038/sdata.2016.134},
	url={https://doi.org/10.1038/sdata.2016.134}
}

@article{piezo-matproj,
	author={de Jong, Maarten
	and Chen, Wei
	and Geerlings, Henry
	and Asta, Mark
	and Persson, Kristin Aslaug},
	title={A database to enable discovery and design of piezoelectric materials},
	journal={Scientific Data},
	year={2015},
	month={Sep},
	day={29},
	volume={2},
	number={1},
	pages={150053},
	issn={2052-4463},
	doi={10.1038/sdata.2015.53},
	url={https://doi.org/10.1038/sdata.2015.53}
}

@Article{elastic-matproj,
	author={de Jong, Maarten
	and Chen, Wei
	and Angsten, Thomas
	and Jain, Anubhav
	and Notestine, Randy
	and Gamst, Anthony
	and Sluiter, Marcel
	and Krishna Ande, Chaitanya
	and van der Zwaag, Sybrand
	and Plata, Jose J.
	and Toher, Cormac
	and Curtarolo, Stefano
	and Ceder, Gerbrand
	and Persson, Kristin A.
	and Asta, Mark},
	title={Charting the complete elastic properties of inorganic crystalline compounds},
	journal={Scientific Data},
	year={2015},
	month={Mar},
	day={17},
	volume={2},
	number={1},
	pages={150009},
	issn={2052-4463},
	doi={10.1038/sdata.2015.9},
	url={https://doi.org/10.1038/sdata.2015.9}
}

@inproceedings{equiformerv2,
	title={{EquiformerV2: Improved Equivariant Transformer for Scaling to Higher-Degree Representations}},
	author={Yi-Lun Liao and Brandon Wood and Abhishek Das and Tess Smidt},
	booktitle={International Conference on Learning Representations (ICLR)},
	year={2024},
	url={https://openreview.net/forum?id=mCOBKZmrzD}
}

@inproceedings{mpnn,
  title={Neural message passing for quantum chemistry},
  author={Gilmer, Justin and Schoenholz, Samuel S and Riley, Patrick F and Vinyals, Oriol and Dahl, George E},
  booktitle={34th Int. Conf. Mach. Learn.},
  pages={1263-1272},
  volume = {70},
  year={2017},
  organization={PMLR}
}

@article{megnet,
  title={Graph networks as a universal machine learning framework for molecules and crystals},
  author={Chen, Chi and Ye, Weike and Zuo, Yunxing and Zheng, Chen and Ong, Shyue Ping},
  journal={Chem. Mater.},
  volume={31},
  number={9},
  pages={pp. 3564--3572},
  year={2019},
  publisher={ACS Publications}
}

@article{chgcnn,
	title={Crystal hypergraph convolutional networks},
	author={Heilman, Alexander J and Gong, Weiyi and Yan, Qimin},
	journal={npj Computational Materials},
	volume={11},
	number={1},
	pages={336},
	year={2025},
	publisher={Nature Publishing Group UK London}
}

@misc{tensor_pred,
	title={Equivariant Graph Neural Networks for Prediction of Tensor Material Properties of Crystals}, 
	author={Alex Heilman and Claire Schlesinger and Qimin Yan},
	year={2024},
	eprint={2406.03563},
	archivePrefix={arXiv},
	primaryClass={physics.comp-ph},
	url={https://arxiv.org/abs/2406.03563}, 
}

@article{gnnopt,
	title={Universal Ensemble-Embedding Graph Neural Network for Direct Prediction of Optical Spectra from Crystal Structures},
	author={Hung, Nguyen Tuan and Okabe, Ryotaro and Chotrattanapituk, Abhijatmedhi and Li, Mingda},
	journal={Advanced Materials},
	volume={36},
	number={46},
	pages={2409175},
	year={2024},
	publisher={Wiley Online Library}
}

@incollection{wyckoff,
	title={Remarks on Wyckoff positions},
	author={M{\"u}ller, Ulrich},
	booktitle={International Tables for Crystallography Volume A1: Symmetry relations between space groups},
	pages={24--26},
	year={2006},
	publisher={Springer}
}

@article{pymatgen,
	title={Python Materials Genomics (pymatgen): A robust, open-source python library for materials analysis},
	author={Ong, Shyue Ping and Richards, William Davidson and Jain, Anubhav and Hautier, Geoffroy and Kocher, Michael and Cholia, Shreyas and Gunter, Dan and Chevrier, Vincent L and Persson, Kristin A and Ceder, Gerbrand},
	journal={Computational Materials Science},
	volume={68},
	pages={314--319},
	year={2013},
	publisher={Elsevier}
}

@inproceedings{canonical_tensor,
	title={Revisiting the Canonicalization for Fast and Accurate Crystal Tensor Property Prediction},
	author={Hua, Haowei and Yang, Jingwen and Lin, Wanyu and Zhou, Pan},
	booktitle={Proceedings of the AAAI Conference on Artificial Intelligence},
	volume={40},
	number={1},
	pages={417--425},
	year={2026}
}

@article{spglib,
	author = {Atsushi Togo and Kohei Shinohara and Isao Tanaka},
	title = {Spglib: a software library for crystal symmetry search},
	journal = {Science and Technology of Advanced Materials: Methods},
	volume = {4},
	number = {1},
	pages = {2384822},
	year = {2024},
	publisher = {Taylor \& Francis},
	doi = {10.1080/27660400.2024.2384822},
	URL = { 
	https://doi.org/10.1080/27660400.2024.2384822
	},
	eprint = { 
	https://doi.org/10.1080/27660400.2024.2384822
	}
}

@article{e3nn,
  title={e3nn: Euclidean neural networks},
  author={Geiger, Mario and Smidt, Tess},
  journal={arXiv preprint arXiv:2207.09453},
  year={2022}
}

@article{tfn,
      title={Tensor field networks: Rotation- and translation-equivariant neural networks for 3D point clouds}, 
      author={Nathaniel Thomas and Tess Smidt and Steven Kearnes and Lusann Yang and Li Li and Kai Kohlhoff and Patrick Riley},
	  journal={arXiv preprint arXiv:1802.08219},
      year={2018},
      eprint={1802.08219},
      archivePrefix={arXiv},
      primaryClass={cs.LG},
      url={https://arxiv.org/abs/1802.08219}, 
}

@article{o3transformer1,
      title={Complete and Efficient Graph Transformers for Crystal Material Property Prediction}, 
      author={Keqiang Yan and Cong Fu and Xiaofeng Qian and Xiaoning Qian and Shuiwang Ji},
      year={2024},
      journal={arXiv preprint arXiv:2403.11857},
      eprint={2403.11857},
      archivePrefix={arXiv},
      primaryClass={cs.LG},
      url={https://arxiv.org/abs/2403.11857}, 
}

@article{mlreview1,
  title={Machine learning in materials science},
  author={Wei, Jing and Chu, Xuan and Sun, Xiang-Yu and Xu, Kun and Deng, Hui-Xiong and Chen, Jigen and Wei, Zhongming and Lei, Ming},
  journal={InfoMat},
  volume={1},
  number={3},
  pages={pp.  338--358},
  year={2019},
  publisher={Wiley Online Library}
}

@article{mlreview2,
title = {Machine learning in materials genome initiative: A review},
journal = {J. Mater. Sci.  Tech.},
volume = {57},
pages = {pp. 113-122},
year = {2020},
issn = {1005-0302},
doi = {https://doi.org/10.1016/j.jmst.2020.01.067},
url = {https://www.sciencedirect.com/science/article/pii/S1005030220303327},
author = {Yingli Liu and Chen Niu and Zhuo Wang and Yong Gan and Yan Zhu and Shuhong Sun and Tao Shen}
}

@article{mlreview3,
title = {Scope of machine learning in materials research—A review},
journal = {Appl. Surf. Sci. Adv.},
volume = {18},
pages = {article 100523},
year = {2023},
issn = {2666-5239},
doi = {https://doi.org/10.1016/j.apsadv.2023.100523},
url = {https://www.sciencedirect.com/science/article/pii/S2666523923001575},
author = {Md Hosne Mobarak and Mariam Akter Mimona and Md. Aminul Islam and Nayem Hossain and Fatema Tuz Zohura and Ibnul Imtiaz and Md Israfil Hossain Rimon},
}

@article{geocgcnn,
  title={A geometric-information-enhanced crystal graph network for predicting properties of materials},
  author={Cheng, Jiucheng and Zhang, Chunkai and Dong, Lifeng},
  journal={Commun. Mater.},
  volume={2},
  number={1},
  pages={article   92},
  year={2021},
  publisher={Nature Publishing Group UK London}
}

@article{icgcnn,
  title={Developing an improved crystal graph convolutional neural network framework for accelerated materials discovery},
  author={Park, Cheol Woo and Wolverton, Chris},
  journal={Phys. Rev. Mater.},
  volume={4},
  number={6},
  pages={article   063801},
  year={2020},
  publisher={APS}
}
	
\end{document}